%% file: REV_1.tex
\documentclass[lettersize,journal]{IEEEtran}

\usepackage{cite}
\usepackage{amsmath,amssymb,amsfonts}
\usepackage{algorithmic}
\usepackage{graphicx}
\usepackage{textcomp}
\usepackage{lipsum}
\usepackage{url}
\usepackage[detect-none]{siunitx}
\usepackage[font=footnotesize]{caption}
\usepackage{tabularx,booktabs}
\usepackage{hhline}
\usepackage{datetime2}
\usepackage{subcaption}
\usepackage{multirow}
\usepackage{eurosym}

\usepackage{epstopdf}

\usepackage{array}
\usepackage{verbatim}
\def\BibTeX{{\rm B\kern-.05em{\sc i\kern-.025em b}\kern-.08em
    T\kern-.1667em\lower.7ex\hbox{E}\kern-.125emX}}
\usepackage{balance}

\def\BibTeX{{\rm B\kern-.05em{\sc i\kern-.025em b}\kern-.08em
    T\kern-.1667em\lower.7ex\hbox{E}\kern-.125emX}}

\begin{document}
\bstctlcite{IEEEexample:BSTcontrol}

\title{Aerosense: A Self-Sustainable And Long-Range Bluetooth Wireless Sensor Node for Aerodynamic and Aeroacoustic Monitoring on Wind Turbines}

\author{Tommaso Polonelli, \IEEEmembership{Member, IEEE}, Hanna M\"uller, \IEEEmembership{Student Member, IEEE}, Weikang Kong, Raphael Fischer, Luca Benini, \IEEEmembership{Fellow, IEEE} and Michele Magno, \IEEEmembership{Senior Member, IEEE}

\thanks{\today, Manuscript created May, 2022. 
This work was supported in part by the Swiss National Science Foundation (SNSF) Bridge Project “AeroSense” under Project 40B2-0\_187087. Moreover, authors sincerely thanks Dr. Julien Deparday for his valuable support as aerodynamic expert in addiction to Fu Zheng for his practical support during the Aerosense testing.
}}


\maketitle

\begin{abstract}
This paper presents a low-power, self-sustainable, and modular wireless sensor node for aerodynamic and acoustic measurements on wind turbines and other industrial structures. It includes 40 high-accuracy barometers, 10 microphones, 5 differential pressure sensors, and implements a lossy and a lossless on-board data compression algorithm to decrease the transmission energy cost. The wireless transmitter is based on Bluetooth Low Energy 5.1 tuned for long-range and high throughput while maintaining adequate per-bit energy efficiency (80~nJ). Moreover, we field-assessed the node capability to collect precise and accurate aerodynamic data. Outdoor experimental tests revealed that the system can acquire and sustain a data rate of 850~kbps over 438~m. The power consumption while collecting and streaming all measured data is  120~mW, enabling self-sustainability and long-term \textit{in-situ} monitoring with a 111~cm\textsuperscript{2} photovoltaic panel. 
\end{abstract}

\begin{IEEEkeywords}
IoT, Wind Turbines, Sensors, Low Power, Energy Harvesting, Structural Health Monitoring, Aerodynamics
\end{IEEEkeywords}

\input{text/intro}
\input{text/relatedworks}
\input{text/background}
\input{text/sys}
\input{text/DataCompression}
\input{text/results}
\input{text/conclusion}


\bibliographystyle{IEEEtran}
\bibliography{bstctl,bibliography}

\end{document}

%% file: text/intro.tex
\section{Introduction}
\label{sec:introduction}
%
Wind energy contributes to climate-neutral energy production and thus mitigates global warming~\cite{suryakiran2020development}. The need to shift away from fossil energy sources has become ever more urgent. In the near future, wind energy must rise from today’s 6\% share of the global power mix to more than 30\% by 2025 to achieve proximity to a pathway well below $2^\circ C$ increase in global warming~\cite{suryakiran2020development}. 
To match the tremendous need for an efficiency increase in wind electricity generation, the energy harvested per square meter must be improved, or wind turbines should be installed on more sites.
This requires a deeper understanding of the blade aerodynamics and a reduced environmental impact. As for sites selection, currently, the installation of wind turbines is constrained due to their significant noise emission and the subsequent complaints from nearby residents~\cite{deshmukh2019wind}. 
It is indispensable to
collect the pressure gradient and audio data from operating wind turbines to better understand aerodynamics and aeroacoustic phenomena. Indeed, since this research field is relatively new, there is a clear need for advanced solutions for continuous monitoring and data collection, especially for producing publicly available datasets~\cite{fischer2021windnode}. 
Supervisory Control and Data Acquisition (SCADA) systems are designed and developed in a wide range of industrial control and the Industrial Internet of Things (IIoT)~\cite{di2021structural}. However, they are only minimally present in the wind turbine industry, and they are not acquiring data from the blades but rather from statics parts \cite{jin2020condition}. This is due to the challenging task of acquiring aerodynamic and aeroacoustic data from the blades of a hundreds meters' tall turbine. In practice, this usually implies considerable effort and cost in measurement campaigns~\cite{jin2020condition,madsen2010dan}.
Recently, some approaches have been proposed to acquire data from the blades showing the importance of such measurements. For instance, in~\cite{mieloszyk2017application}, the authors manufactured an entire custom blade, with an embedded Pitot tube and acoustic sensors~\cite{mieloszyk2017application,aagaard2010dan}, to measure the airstream and turbulences impacting the rotor blade. However, a recent trend is to replace such expensive and wired SCADA systems with scalable and low-power IIoT sensors for the next generation of monitoring systems~\cite{fischer2021windnode}, characterized by a limited installation cost. 
%

%

Although many SCADA solutions are moving to the IIoT field for both the Structural Health Monitoring (SHM) and aerodynamic blade monitoring, today, they are still bulky, expensive, and require to be powered by the wind turbine itself, which often makes this deployment impossible~\cite{maldonado2020using}. The most attractive solutions are represented by small size, battery-operated IoT nodes connected to the cloud with a wireless interface~\cite{ahmad2021scopes}. However, when a small form factor and long battery lifetime~\cite{di2021structural} are required, there are still unsolved design challenges to be addressed. Some of those are the energy consumption in active and sleep mode, the communication power, range and bandwidth~\cite{ballerini2020nb} - all critical parameters in this application scenario due to the size of the turbine and the number of deployed sensors~\cite{fischer2021windnode}.
%
Thus, it is crucial to design energy-efficient and reliable devices able to acquire, process and transmit data for several years and, eventually, for the full wind turbine lifetime, exploiting environmental sources such as small solar photovoltaic panels~\cite{fischer2021windnode,di2021structural,ahmad2021scopes}.

This paper presents a self-sustaining monitoring system for studying the aerodynamics of wind turbines with multi Micro-ElectroMechanical Systems (MEMS) sensors for the next generation of SCADA devices. The proposed system is called Aerosense, and features arrays of pressure and acoustic MEMS sensors and an Inertial Measurement Unit (IMU) coupled with edge computing solutions and photovoltaic energy harvesting to achieve in-field precise and sustainable data collection. 
In particular, it is suitable for installation on a wide range of wind turbine blades from a few meters to hundreds of meters, hosting a modular sensor unit with up to 40 barometers and 10 microphones, plus 5 differential pressure sensors. The wireless communication is based on Bluetooth Low Energy 5.1 (BLE). 
It employs a power amplifier featuring long-range, up to 400~m, coupled with $\sim$1~Mbps bandwidth to stream all the collected data.
Aerosense also features on-board data processing for reducing the volume of data transmitted: we developed two different compression algorithms: a lossless inter-channel redundancy removal for barometers and a lossy spectrum compression code for microphones.
The designed node has been tested and evaluated in-field regarding functionality, energy consumption, and range.
We determined that the sensor node reaches self-sustainability with a $111~cm^2$ photovoltaic panel, or more than 100 days in complete darkness conditions using only the battery. 
The main contributions of this paper are summarized as follows:
(\textit{i}) To present the architecture, the design and the evaluation of a self-sustaining multi-sensor wireless platform that enables \textit{in-situ} and long-term monitoring on a heterogeneous variety of wind turbines.
(\textit{ii}) To prove the feasibility of acquiring aerodynamic and aeroacoustic data using a modular and low-power IoT device and the wireless connection with the cloud. 
(\textit{iii}) To present an advanced power management circuit with a multi-cluster power supply and photovoltaic energy harvesting to achieve self-sustainability.
(\textit{iv}) To propose two specific compression algorithms targeted for audio and pressure data, respectively, with lossy and lossless encoding.
(\textit{v}) To experimentally evaluate the designed sensor node in terms of functionality, energy efficiency of the compression algorithm, energy consumption, and lifetime. 
%

%
Aerosense is a self-sustainable and modular IoT solution that can monitor and collect aerodynamic and aeroacoustic information on operating wind turbines. It enables a future generation of real-time and predictive maintenance devices for optimizing renewable energy production. 

%% file: text/relatedworks.tex
\section{Related Works}
\label{sec:related}
Monitoring civil structures and wind turbines, i.e., structural health monitoring, with IoT systems has been a very active research topic in recent years~\cite{aagaard2010dan,fischer2021windnode,di2021structural,ballerini2020nb}. In previous works, wind turbines have been equipped with pressure sensors~\cite{madsen2010dan,aagaard2010dan}, and microphones~\cite{fischer2021windnode}, demonstrating their importance for real-time control and smart maintenance. 
Although previous analysis~\cite{fathima2021mems,di2021structural} has shown the potential of using inexpensive MEMS sensors for SHM applications, other than our work, no MEMS multi-sensor array for wind turbines is present in the literature, nor in commercial products. Arrays of barometric MEMS sensors have already been studied in several other application scenarios, where arrays of pressure sensors have been mounted onto aerodynamic surfaces, namely on airplane wings and cars~\cite{filipsky2017design}. In general, the aforementioned works show that MEMS sensors can be used to acquire aerodynamic and aeroacoustic measurements on surfaces. However, they do not address the wireless communication and power consumption challenges and requirements of continuous and long-term monitoring of wind turbines~\cite{karad2021efficient}.
Moreover, in one of our previous papers~\cite{fischer2021windnode}, we demonstrated the practical possibility to use MEMS sensors in this context, such as measuring the sound level and the pressure distribution on a blade. 
In the literature, there are already a few examples of wireless solutions installed on wind turbines~\cite{wondra2019wireless,di2021structural,lu2019wind}. However, previous designs mainly included vibration measurements~\cite{di2021structural, esu2016feasibility} on the wind turbine tower for SHM monitoring purposes rather than proposing a solution capable of acquiring aerodynamic data directly on the blade(s). In these scenarios, the system has to process and transmit an amount of data in the range of 5~kbps, while for aerodynamic analysis, a throughput larger than 1~Mbps is necessary~\cite{fischer2021windnode}. The additional data collected by arrays of barometers and microphones is crucial for wind turbine modeling, but on the other hand, it poses major challenges in the design of energy-efficient and long-lasting IoT devices.
%
%

%
The authors in~\cite{wondra2019wireless} propose a flexible solution offering a  battery-powered and wireless system for SHM of wind turbine towers exploiting only MEMS accelerometers. Despite the authors deploying the system on a wind turbine, they do not provide any data about the power consumption and estimated battery lifetime. 
Another comparable solution is proposed in~\cite{lu2019wind}, where the health status of a planetary gearbox is estimated through vibration analysis. This work also shows the practical implementation of a vibration-based energy harvester, using a 0.9~mW piezoelectric element to cover the whole energy demand. However, the sensor node features a low sampling frequency of 50~Hz and limited coverage of 6~m. Another SHM platform measuring vibration is proposed in~\cite{di2021structural}, where Di Nuzzo et al. overcome the communication range issue using both LoRaWAN and NB-IoT protocols but supporting only a few kbps of transmission bandwidth. 
%
%

%
Previous works, like~\cite{di2021structural} and~\cite{ballerini2020nb} show that wireless connectivity is one of the most relevant sources of consumption for long-range coverage. In~\cite{di2021structural}, the energy per bit reaches $42~\mu J$ exploiting the cellular infrastructure, while in~\cite{ballerini2020nb} the multi-kilometers coverage is reached at the cost of milli-Joules per bit. These results show that decreasing the data to be transmitted further increases the operation lifetime of a battery-supplied device. Thus, due to the large amount of data handled by our Aerosense, e.g., $200\times$ higher than~\cite{di2021structural}, on-board data compression is an essential feature for our scenario.
%

%
%
%

%
To the best of our knowledge, Aerosense is the first self-sustaining, MEMS-based, long-range BLE system for aerodynamic and aeroacoustic analysis on wind turbine blades, supporting heterogeneous and high-frequency sensors. 

%% file: text/background.tex
\section{Sensing requirements for wind turbine aerodynamics and aeroacoustics}
\label{sec:background}
\begin{figure}[t]
\begin{center}
\includegraphics[width=0.95\columnwidth]{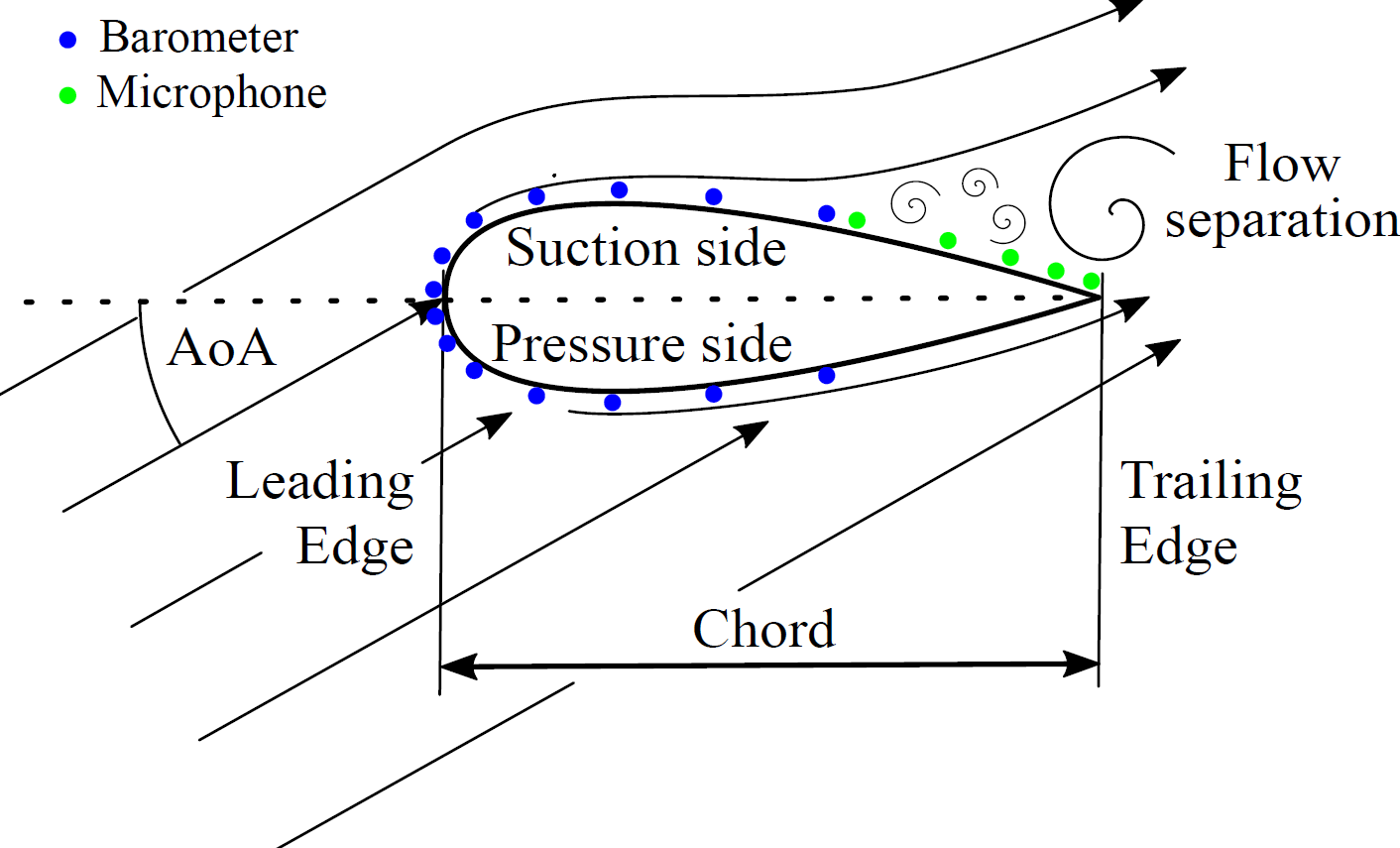}
\caption{Wind turbine blade: airfoil section showing the unwanted flow separation near the Trailing Edge (TE) and the Angle of Attack (AoA)}
\label{fig:airfoil}
\end{center}
\end{figure}

%
The cross-section of a wind turbine blade is called an airfoil. 
For simplicity, a symmetrically shaped airfoil is considered, where the \textit{chord}\footnote{The distance between Leading Edge (LE) and Trailing Edge (TE)} distance is shown in Fig.~\ref{fig:airfoil}.
One of the most critical parameters of a blade is the Angle of Attack (AoA), which affects the angle ($\alpha$) between the chord and the flow direction of the fluid flow where the blade is rotating. An ideal symmetric airfoil does not generate any lift for AoA equal to zero; however, in the $0-20^{\circ}$ range, the pressure on the wind-facing side (Pressure side in Fig.~\ref{fig:airfoil}) increases while it decreases on the suction side, at the opposite edge. In this condition, the pressure distribution surrounding the blade creates the lift force, pulling the blade towards the suction side. When the AoA increases too much, the fluid flow that normally moves over the suction surface generates a flow separation near the TE, causing turbulence, audible noise and dramatically decreasing the lift effect. 
In~\cite{madsen2010dan}, the authors developed and demonstrated an identification method for transitional and turbulence flows relying on 2048 data points, which corresponds to
a sampling time of 0.041 seconds, and a frequency range of 2000-6000~Hz. Moreover, a more recent work~\cite{madsen2019transition} improved the transition detection accuracy by enlarging the audio bandwidth up to 7~kHz. Reports in~\cite{aagaard2010dan,madsen2010dan} show the maximum estimated acoustic pressure level is around $500~Pa^2$, which corresponds to a Sound Pressure Level (SPL) above 140~dB~SPL in the loudest measurement. 
%

%
Moreover, the needed accuracy $P_{acc,min}$ of the absolute pressure sensors was defined to be at most 1\% of the dynamic pressure range $P_{dyn,avg}$ at average wind speeds~\cite{fischer2021windnode}.
$P_{acc,min}$ was determined for wind turbines of three different sizes, Aventa AV-7 turbine~\cite{aventa2021},  DTU 10MW wind turbine~\cite{bak2013dtu}, NREL 5~MW~\cite{jonkman2009definition}, in the range between $5~Pa$ and $40~Pa$ with a $P_{dyn,avg}$ above 1~kPa.
Additionally, the chosen sensor must have a low drift in temperature and aging to avoid recurrent re-calibrations once mounted on a wind turbine. For the sampling rate, in~\cite{fischer2021windnode} it was defined that a sampling of 100 times per rotation generated high-quality measurements and model reconstruction. This amounts to approximately 70 samples per second for the Aventa AV-7 turbine~\cite{aventa2021} and 17~sps for the DTU 10~MW generator~\cite{bak2013dtu}.

%% file: text/sys.tex
\section{System Architecture}
\label{sec:sys}
\begin{figure}[t]
\begin{center}
\includegraphics[width=0.95\columnwidth]{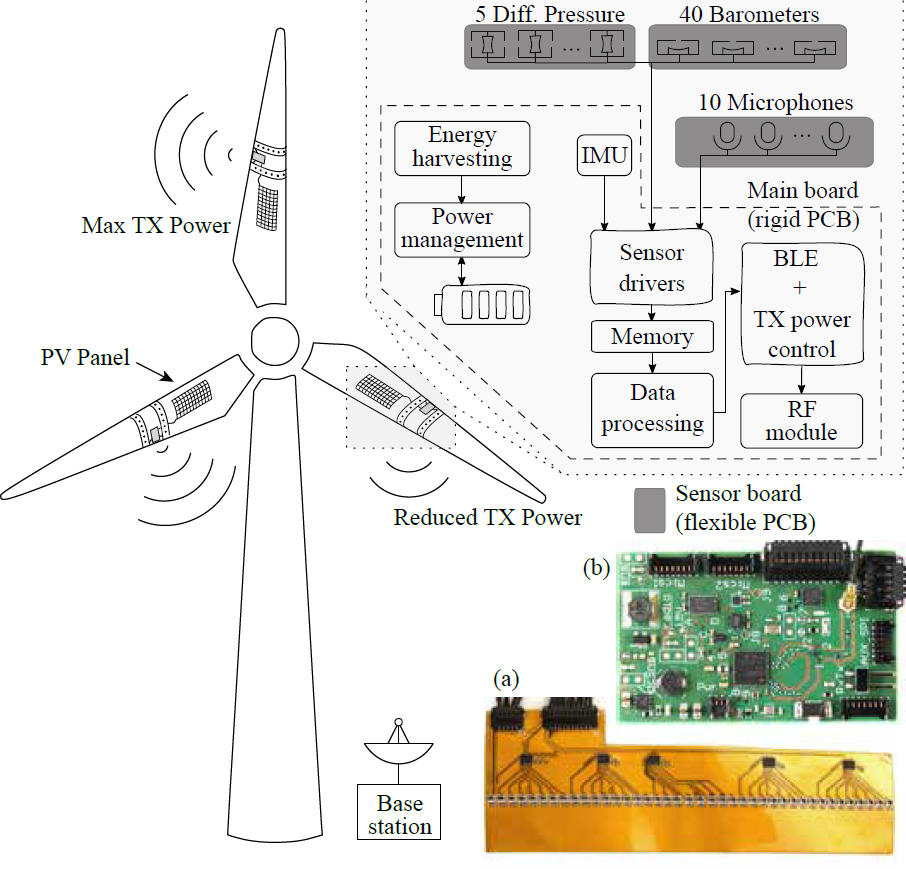}
\caption{A high-level overview of the proposed monitoring system and the Aerosense sensor board. (a) Picture of the 40 barometers flexible PCB tested in the field. (b) View of the main board}
\label{fig:everything}
\end{center}
\end{figure}
This section presents the architecture of the proposed self-sustaining sensor node, called Aerosense. Fig.~\ref{fig:everything} shows the high-level functionality and an intended deployment of the Aerosense sensor node, together with its position on the blade. The energy harvesting part, a vertically oriented flexible solar cell of 111~cm\textsuperscript{2} size and a rechargeable battery with 32~Wh (8.7~Ah), is evaluated, which allows the sensor to work perpetually within specific environmental conditions. For this reason, all the hardware and firmware have been designed to satisfy the project requirements at the lowest energy cost. The sensor selection is based on the aeroacoustic specifications provided in Section~\ref{sec:background}.  
%
\subsection{Sensors}
%
%

%
Following Section~\ref{sec:background}, the pressure difference among the airfoil needs to be measured. Hence it was decided to use an array of 40 absolute pressure sensors at a predefined distance from the trailing edge (see Fig.~\ref{fig:airfoil}) to determine the pressure differences computationally. In addition, Fig.~\ref{fig:airfoil} illustrates the placement of the barometers, in blue, and the microphones, in green, around the blade. 
In Table~\ref{table:barometers}, the key parameters of the best-in-class MEMS barometers are compared.
The targeted absolute accuracy of 5~Pa~\cite{aventa2021} is not available in any commercial production. However, this can be addressed by applying a custom calibration, which can improve the accuracy substantially~\cite{fischer2021windnode}. In Table~\ref{table:barometers}, the LPS27HHW is the one with the best absolute accuracy. Indeed, it sacrifices the power consumption, which is $6.4\times$ higher than the MS5840-02BA to improve the performance. On the other hand, the BMP390L cannot be selected despite its promising features due to a non-water-resistant package, and the PBM230 guarantees an absolute and relative accuracy far from our specifications. 
Finally, the ST LPS27HHW was chosen because it is the best-performing water-resistant sensor, and it also includes a temperature sensor to enable custom and on-site calibrations.
The LPS27HHW is an ultra-compact piezoresistive barometer. 
It can measure from 260 to 1260 hPa absolute pressure range, which covers the range for wind turbine monitoring~\cite{aagaard2010dan,madsen2010dan,aventa2021,bak2013dtu}. 
\begin{table}[t]
\centering
\caption{MEMS sensors comparison table for the Aerosense hardware design.}
\label{table:barometers}
\scriptsize
\setlength{\tabcolsep}{3pt}
\begin{tabular}{l c c c c }
\hline\hline
\multirow{2}{*}{\textbf{Model}} & \textbf{Water} & \textbf{Min. power} & \textbf{Relative/Absolute}  & \textbf{Height} \\
 &  \textbf{resistance} & \textbf{[$\mu W$]} & \textbf{accuracy} [Pa] &  [mm] \\
\hline
Pewatron PBM230	& yes & 5.1			& $\pm$100/$\pm$150 & 2.6 \\ 
ST LPS27HHW 	& yes	& 7.2			& $\pm$2.5/$\pm$100 & 1.7 \\ 
TE MS5840-02BA 	& yes	& 1.13			& $\pm$100/ $\pm$200 & 1.7 \\ 
Bosch BMP390L	& no	& 5.61			& $\pm$3/$\pm$50 & 0.75\\
\hline
Pewatron 52 & yes & 7.3 & $\pm$10/$\pm$50 & 5\\ 
\hline\hline
\multirow{2}{*}{\textbf{Model}}	& \textbf{AOP$^\star$}	& \textbf{Bandwdith} & \textbf{Power}  & \textbf{Height} \\
 & [dB] & [kHz] & [$\mu W$] & [mm] \\
\hline
Vesper VM2020  & 152 & 8 & 446 & 1.3 \\
TDK	ICS-40638 & 138 & 10 & 306 & 1 \\
TDK	T5818 & 135 & 10 & 1062 & 1 \\
ST	MP23DB01HP & 135 & 10 & 1440 & 1.1 \\ 
\hline\hline
\multicolumn{5}{l}{$^\star$ Acoustic Overload Point}
\end{tabular}
\end{table}
Aerosense also supports five differential pressure sensors to open the possibility of true-differential measurement. We selected the Pewatron 52-series 
(see Table~\ref{table:barometers}) due to its good accuracy and stability. Moreover, it offers a compact solution with only 5~mm thickness and a multi-order compensation for offset correction, sensitivity, thermal errors and a non-linearity correction algorithm with calibration coefficients stored in on-chip EEPROM.
%
The sampling rate is hard-coded at 1.2~kHz with 16-bits resolution.
As described in Section~\ref{sec:background}, the recorded audio spectrum is crucial to detect the flow separation effect on the wind turbine. Since microphones generate a large amount of data that is hardly managed by a low-power MCU, we selected the minimum recommendations, including a 6~kHz bandwidth and a maximum pressure level of 140~dB SPL. In Table~\ref{table:barometers}, a list of possible commercial solutions is presented. However, few microphones feature an acoustic overload point above 140~dB, limiting the design choice to only two components, the Vesper VM2020 and the TDK ICS-40638. Despite, as seen in Table~\ref{table:barometers}, the latter has a 20\% larger bandwidth and a 30\% lower power consumption; during in-field comparison, we noticed signal clippings due to audio spikes above 120~dB SPL, and, thus, an undesirable spectrum distortion. The Vesper VM2020 was selected for the Aerosense design, supported by the Texas Instruments ADS131M06 for the analog to digital conversion (ADC).
%

%
The last MEMS sensor on the Aerosense is an IMU (Inertial Measurement Unit). Indeed, vibration analysis on wind turbine blades is an important parameter in assessing their structural health~\cite{di2021structural}. Moreover, it is essential to estimate and correlate the blade position together with pressure and audio data. 
The two 9-axis IMUs considered were the Bosch BMX160 and the TDK InvenSense ICM-20948. For the Aerosense, the former was chosen for its lower active current consumption ($46~\mu A$ vs. $70~\mu A$ for a sampling rate of 100~Hz) and the lower idle current consumption ($4~\mu A$ vs. $8~\mu A$).

The Aerosense measurement system features a maximum thickness of \SI{4}{\milli\metre} and an average of \SI{3}{\milli\metre}, including the battery, connectors, and cabling. In Fig.~\ref{fig:everything} - (a), the flexible PCB is \SI{1.85}{\milli\metre} thick, orders of magnitude lower than a standard blade chord length - Fig.~\ref{fig:airfoil}.  Our experimental analysis did not notice any effect on the blade's aerodynamic performance and structural integrity.
\subsection{Wireless System on Chip}
The wireless interface, including a microcontroller for on-board data processing, and a non-volatile memory chip (Kioxia Flash TC58CYG2S0HRAIJ) of 512~MB have been selected by determining the required transmission range with minimal power consumption, overview in Fig.~\ref{fig:everything}. In addition to compression, the node will also have to support the raw sensor data storage. The 10 microphones at 16~kHz sample rate at 24-bit per sample result in a bandwidth of 3.8~Mbps. In addition, 40 barometers at 100~Hz @ 24-bits generate a bandwidth of 96~kbps. The data from the IMU and the differential pressure sensors (96~kbps) results in a total required bandwidth of 4.2~Mbps that the MCU needs to handle from the source to the flash, and then from the internal memory to the base station.
Kioxia Flash was then selected since it can support up to 15 minutes of continuous recording, a period comparable with Aerosense project requirements.
Among many different SoCs that embed a wireless BLE transceiver with an ARM Cortex, we selected the Texas Instruments CC2652P. It features an integrated power amplifier enabling long ranges and the possibility to handle the multi-Mbps data flow. Moreover, the CC2652P is designed for energy efficiency when reading out sensors, hosting a Sensor Controller (SC), which enables the use of one line of SPI and I2C at only $30~\mu A$ in addition to standard peripherals supported by the main core. The CC2652P encompasses a 48-MHz ARM Cortex-M4F processor, with 352~kB in-system programmable flash and 88~kB SRAM.
\subsection{Real-Time Transmission Power Control}
\label{sec:RTPC}
Since power consumption is one of the Aerosense key parameters, Aerosense implements a Real-Time Transmission Power Control (RTPC) to ensure no energy is wasted by transmitting at unnecessarily high TX power. Indeed, the distance between the base station and the antenna on the blade can change significantly, i.e., by 80~m~\cite{aagaard2010dan,madsen2010dan}, depending on the relative position of the blade.
As noted already by O. Esu et al.~\cite{esu2016feasibility}, when the distance between transmitter and receiver is not constant, an RTPC algorithm can significantly reduce power consumption by always keeping the TX power at the optimal point required for the current distance.
%
%

First, since there is no power limitation on the base station, its TX power is fixed to 20~dBm. Due to the transmission channel reciprocity in a Line Of Sight (LOS) condition, Aerosense can calculate the Received Signal Strength (RSSI) of the base station with Eq.~\ref{eq:TPC_RSSI}, eliminating the need for control packets, therefore reducing it to a feed-forward controller. The smallest RX power that offers reliable communications was determined experimentally between -64~dBm and -68~dBm. Furthermore, a second-order IIR filter was applied to the RSSI readings. A band pass-frequency of 1~Hz was chosen such that typical distance oscillations on wind turbines are not suppressed.
\begin{equation}
\footnotesize
RSSI_\text{BaseStation} = RSSI_\text{AeroSense} + P_{\text{TX,AeroSense}} - 20~dBm
\label{eq:TPC_RSSI}
\end{equation}
%
In our implementation, the receiver power TX can get raised by a non-negative boost variable. This boost variable is increased under the following conditions and it is summed to the smallest RX power:
\textit{(i)} The base station continuously checks the Packet Error Rate (PER), and if it exceeds 10\%, it sends a command to the Aerosense to increase the boost.
\textit{(ii)} The transmission queue has grown over a certain threshold. 
Note two more implementation details:
    \textit{($\alpha$)} If 100~ms after issuing a boost, the transmission queue is still over the threshold, the boost variable is increased again.
    \textit{($\beta$)} A second higher threshold exists, which, when exceeded, triggers a more significant boost increase. This serves as a safety measure to prevent the buffer from overflowing.
\textit{(iii)} The boost variable is decreased if the base station PER has been below 5\% during six connection events or if the transmission queue sinks below a certain threshold.
\subsection{Power Supply and Energy Harvesting}
The Aerosense features a flexible power supply powered by lithium-ion batteries and an Energy Harvesting (EH) circuit with a dedicated power manager. Thus, it can be used reliably over short (months) periods with
the battery, which is then recharged through the collected environmental energy, potentially self-sustaining the device operation.
Three high-efficiency DC/DC converters (Texas Instruments TPS62842DGRR) were connected directly to the battery to achieve this flexibility since they can convert a wide range of input voltages to a fixed target voltage. To optimize the power consumption, the circuit was split up into six voltage domains. The sensors are driven by their minimum voltage, 1.8~V, and the MCU runs a 3.3~V to enable the maximum radio transmission power. These voltage domains can also be turned off entirely when the system is idle.
%

%
The solar EH relies on two main components, the flexible solar panel MPT3.6-150 from FlexSolarCells\footnote{\scriptsize \url{https://flexsolarcells.com/}} and the Texas Instruments BQ25570. The MPT3.6-150 generates 360~mW in a $111~cm^2$ area, and due to its properties, easily fits on the twisted wind turbine blade. An ultra-low thickness of only 0.2~mm generates negligible effects on the airfoil aerodynamic. The BQ25570 is an ultra-low-power DC-DC boost charger capable of collecting energy from sources as low as 100~mV with an average efficiency above 80\%. As shown in Fig.~\ref{fig:everything}, it does not directly supply the circuitry, but it is directly connected to the battery.  

%% file: text/DataCompression.tex
\section{Data Compression}
\label{sec:compression}
In the Aerosense system architecture, the raw data is collected by the sensors and directly saved on the Flash. Then, the compression algorithm decreases the raw data size, and the CC2652P sends compressed packets to the base station over BLE for further processing and long-term storage.
Audio compression techniques have been extensively studied in the literature~\cite{chowdhury2020adaptive}, but they are not plug-and-play suitable for the application scenario of our interest. Thus, in this section, different existing compression methods are analyzed and implemented on the Aerosense.
More than $160\cdot10^3$ audio data samples are collected every second, yet for pressure data, it is only 4,000. Due to having only a 2.5\% impact on the data traffic, only lossless compression is considered for the latter. For the Pewatron sensors, we applied the same methodology used for the LPS27HHW, also considering that system requirements do not evidence any unnecessary or redundant information on the signal spectrum. 
Pressure data, and other environmental data such as temperature and humidity, are usually considered temporally smooth and vary only in a small range. In~\cite{du2017compressed,chowdhury2020adaptive}, the experiments testing multiple existing compression methods are conducted on typical environmental data on a microcontroller and achieve satisfactory compression ratios together with efficient energy usage. However, the wind turbine pressure data are collected from a rotating wind turbine. The rotation and the wind blow induce less temporally smooth pressure data with a larger dynamic range. This generates a drop in performances for lossless algorithms.
Due to the limited resources on the ARM Cortex M4F processor, the implementation of compression methods is much more constrained than that on a high-performance counterpart. The Aerosense objective is to reduce the size of transmission files in order to save transmission energy consumption. Hence the data compression is meaningful only when the condition in Eq.~\ref{eq:design} is valid. 
\begin{equation}
\footnotesize
{P\sb{cpr}\times\frac{S\sb{orig}}{\Theta\sb{cpr}}+P\sb{tx}\times\frac{S\sb{orig}}{CR\times\Theta\sb{tx}}}\leq{P\sb{tx}\times\frac{S\sb{orig}}{\Theta\sb{tx}}}.
\label{eq:design}
\end{equation}
Where $S_{orig}$ denotes the original file size, $CR$ denotes the compression ratio,  $\Theta_{cpr}$ denotes the throughput of the compression method,  $\Theta_{tx}$ denotes the throughput of BLE data transmission, $P_{cpr}$ denotes the power consumption of the compression method, $P_{tx}$ denotes the power consumption of the data transmission.
\subsection{Lossless Compression for Pressure Data}
In this section, our lossless compression approach for pressure data is described. 
%

%
For multi-channel configurations, e.g., the Aerosense setup with up to 40 parallel sensors, the data from adjacent channels typically feature similarities. These similarities can be utilized for redundancy removal, for example, by subtraction. 
However, performing inter-channel redundancy removal does not consistently smooth the samples. Thus, it is sensible to estimate the necessity of applying it to actual data. One heuristic is to perform inter-channel redundancy removal only when the summation of absolute values of discrete derivatives drops.
In this manuscript, fixed-order predictors are considered and evaluated. 
We noticed that the first-order prediction residuals of different channels are much smaller than those of original channel samples. Besides, it is also clear that performing inter-channel redundancy removal and prediction together offer a significant improvement rather than performing either of them alone.
%
%
%
By investigating the best log-likelihood distribution of the residual, we found that the two most likely estimators are a negative binomial and geometric distribution. The likelihoods of both distributions are similar; thus, we consider the rice coding~\cite{solano2020optimal} a suitable entropy encoder for the residuals. See Eq.~\ref{eq:rice}, where M is the rice parameter, $f$ is the estimation of remainder as a constant fraction $f \in  [0,1]$ of $2^M - 1$, and $\mu$ is the mean value. 
\begin{equation}
\footnotesize
\hat{M}_f(\mu) = max\left \{ 0,\left [ log_2 (\mu +f) \right ] \right \}
\label{eq:rice}
\end{equation}
\subsection{Lossy Compression for Audio Data}
In this chapter, two lossy compression methods for audio data are considered. IMA ADPCM (Interactive Multimedia Association \& Differential Pulse-code Modulation) works in the time domain and is introduced as the comparison baseline. Fast Fourier Transform - High Pass Filter (FFT-HPF) is the proposed frequency-domain compression method. 
%
The implementation in this project is based on the IMA ADPCM implementation provided in~\cite{liaghati2018microcontroller}.
The algorithm behavior could be tuned by adjusting the index change table and step size table, which involves predicting the future residuals. In this project, IMA ADPCM sticks to the original tables from~\cite{liaghati2018microcontroller}.
In the current application scenario, we are only interested in the frequency band higher than 100 Hz. This means that only partial frequency coefficients need to be transmitted after performing FFT on the time-domain data. The transmitted data size can be further reduced by quantizing the coefficients and utilizing the symmetry of FFT coefficients. 
%
%
Thus, a higher compression ratio can be achieved by converting them into integers, or by quantizing them even further. Quantizing all the coefficients with a constant step size will incur high quantization errors. This project proposes to divide frequency coefficients into several smaller bands and determine a step size for each band. The quantization step size is determined according to the maximum magnitude of the frequency coefficients in that band and the number of quantization levels. The desired compression ratio determines the number of quantization levels.
Assuming the original bit depth of audio samples are $D$ bits, the desired compression ratio is $CR$, and the maximum magnitude is $mag = max(abs(s))$, where $s$ is the frequency coefficient vector. The sample mean is not subtracted from the input vector to make it zero-centered because the sample mean is usually small, which means performing the subtraction contributes little to the reduction in quantization error while it leads to
more overheads. To roughly achieve the desired compression ratio, the compressed data should be represented with $d = \frac{D}{CR}$ bits. The number of quantization levels is $l = 2^d$. The
quantization step size is then $\frac{2 \cdot mag}{l}$. Note that the quantization step size for each band is stored in a table and that both the coefficients and the table are needed to decompress the data.
Lastly, it is found that the quantization step sizes for adjacent bands/channels are similar and sometimes they are the same. Then, first calculating the difference of adjacent quantization step sizes and subsequently performing the run-length encoding for consecutive zeros is beneficial to increase the compression ratio.

%% file: text/results.tex
\section{Experimental Results}
\label{sec:results}
%
In the final prototype version, Aerosense comprises four separate elements, three flexible PCBs hosting sensors and the main sensor board that hosts the system on chip with Bluetooth 5.1, the power management, and the energy harvester.
The sensor PCBs are designed in Kapton flexible substrate, allowing them to be wrapped around the blade. Moreover, this technical choice allows a lower thickness than the bound enforced by the aerodynamics non-perturbation constraint - \SI{4}{\milli\metre}.
In this work, to show the performance in the worst-case scenario, we tested the prototype with 10 microphones, 40 barometers, and 5 differential pressure sensors.
\subsection{Wireless Communication}
We tested the system in static conditions to verify the communication link and also on an Aventa AV-7 wind turbine~\cite{aventa2021}.
In static conditions, the BLE transfers data reliably up to 275~m with an average bitrate of 938~kbps. Furthermore, the system reaches a maximum of 438~m with a decreased speed of 850~kbps due to packet loss. 
The highest measured bandwidth was 1.2~Mbps at the distance of 200~m, which satisfies the application specification listed in Section~\ref{sec:background}.
%

%
%
On average, we measured that the BLE energy per bit is 80~nJ/byte and 180~nJ/byte, respectively, for small (Aventa AV-7~\cite{aventa2021}) and large (10~MW~\cite{bak2013dtu}) wind turbines, confirming the effective RTPC energy reduction.
\subsection{Lossy and Lossless: In-Field Results}
\begin{figure}
    \centering
    \begin{subfigure}[t]{0.48\columnwidth}
        \centering
        \includegraphics[width=\columnwidth]{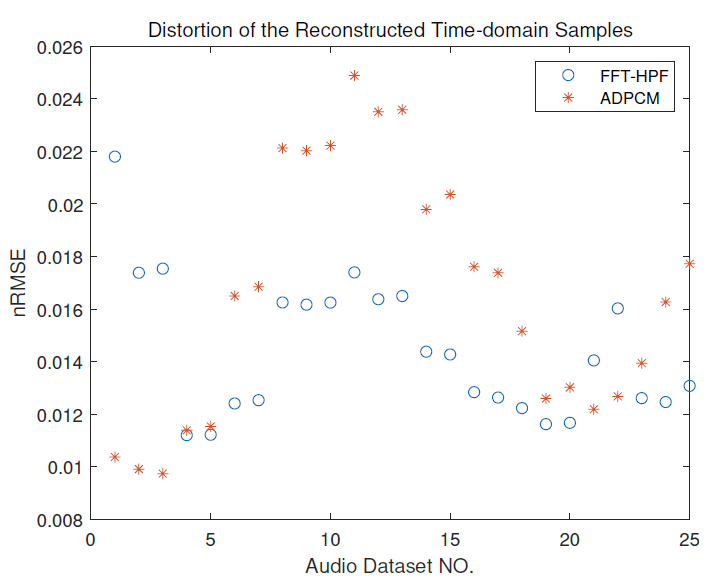}
        \caption{nRMSE}
        \label{fig:rmse}
    \end{subfigure}
    \hfill
    \begin{subfigure}[t]{0.48\columnwidth}
        \centering
        \includegraphics[width=\columnwidth]{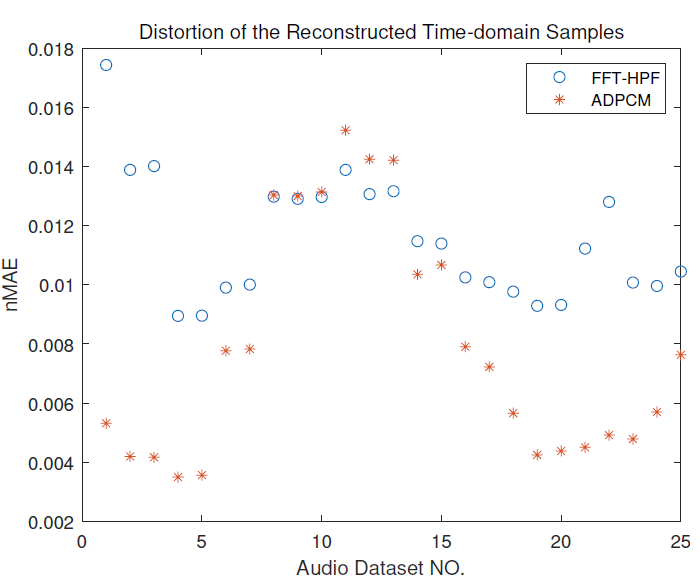}
        \caption{nMAE}
        \label{fig:mae}
    \end{subfigure}
    \\
    \begin{subfigure}[b]{0.48\columnwidth}
        \centering
        \includegraphics[width=\columnwidth]{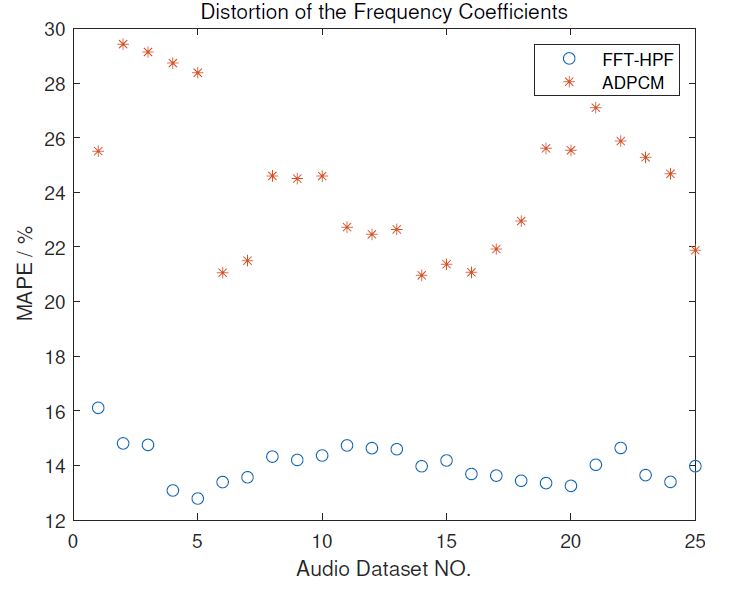}
        \caption{MAPE of Frequency Coefficients}
        \label{fig:mape}
    \end{subfigure}
    \hfill
    \begin{subfigure}[b]{0.48\columnwidth}
        \centering
        \includegraphics[width=\columnwidth]{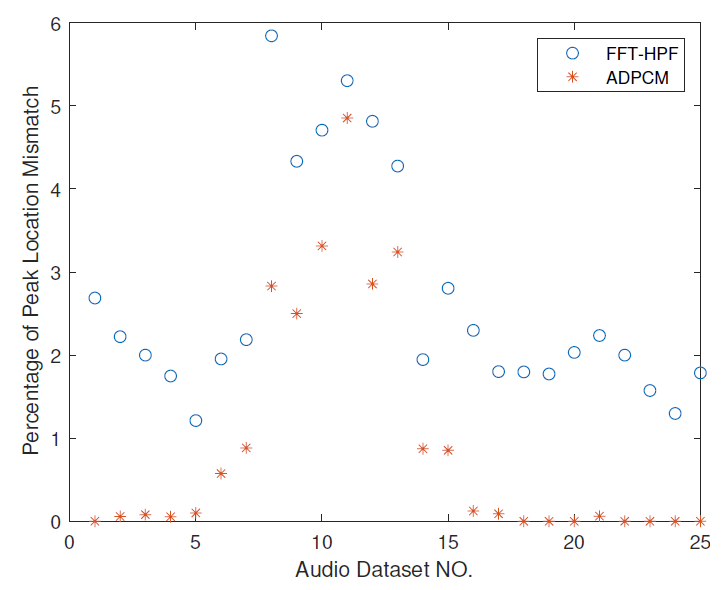}
        \caption{Percentage of Peak Location Mismatch}
        \label{fig:mismatch}
    \end{subfigure}
    \caption{Distortion of audio Samples for FFT-HPF and ADPCM lossy compression algorithms among different deployments (datasets).}
    \label{fig:distortion}
\end{figure}

To evaluate the performance of the proposed methods, it is critical to know the compression ratio and the energy consumption per byte. For lossy data
compression, since only partial information is kept, the remaining information should reflect the raw signal without much distortion. This distortion is measured with generic metrics such as Mean Absolute Error (MAE), Root Mean Square Deviation (RMSE), and Mean Absolute Percentage Error (MAPE). Furthermore, since the decompressed data is used for analytical purposes, the following application-specific metric is considered: percentage of peak location mismatch.
The methods proposed in Section~\ref{sec:compression} are evaluated on pressure and audio data. Each audio and pressure block has
1024 and 512 entries, respectively.
Table~\ref{table:lossless} shows experiment results for lossless compression on pressure data. The achieved mean compression ratio ($\mu CR$) is limited to a factor of two due to the large dynamic range of the pressure distribution, and in the worst case, it decreases to 1.28. However, when processing the data on-board, the overall energy consumption decreases by 46\% for sending raw data to the base station.
\begin{table}[t]
\centering
\caption{Comparison of Lossless and Lossy Compression Methods for Pressure and Audio Data. Percentage of Energy Saving (PES). Break-even Compression Ratio (BECR)}
\label{table:lossless}
\setlength{\tabcolsep}{3pt}
\begin{tabular}{l c c c}
\hline\hline
 & \multirow{2}{*}{\textbf{Pressure Data}} & \multicolumn{2}{c}{\textbf{Audio Data}} \\
 & & FFT-HPF & ADPCM \\
\hline
Mean Compression Ratio ($\mu CR$) 	& 2.12 & 4.024 & 4.0 \\ BECR & 1.28 & 1.46 & 1.35\\ 
PES small turbine 	& 30.9\% & 43.9\% & 49.0\%\\
PES large turbine & 46.1\% & 65.7\% & 67.1\% \\
\hline\hline
\end{tabular}
\end{table}
For audio processing, proposed methods in Section~\ref{sec:compression} are evaluated. The compression ratio for ADPCM and FFT-HPF is set to 4.000 since it was experimentally found as a good compromise between energy-saving and reconstruction fidelity. However, the achieved compression is slightly higher because of the coefficient selection. The mean FFT-HPF compression ratio achieved in our implementation is 4.024.
%
The generated signal distortion is shown in Fig.~\ref{fig:distortion}. In Fig.~\ref{fig:rmse} and Fig.~\ref{fig:mae}, FFT-HPF generally has higher MAE but lower RMSE than ADPCM. Considering that RMSE penalizes large errors, the comparison result suggests more large deviations in reconstructed ADPCM time-domain samples.
The distortion of the magnitude of frequency coefficients is evaluated with MAPE. Original and reconstructed ADPCM samples are filtered after being transformed into the frequency domain for the fairness of comparison. The comparison result is shown in Fig.~\ref{fig:mape}, where FFT-HPF performs $+10\%$ better on average. 
Fig.~\ref{fig:mismatch} shows the percentage of peak location mismatch, an application-specific metric. ADPCM performs $2\%$ better than FFT-HPF on average. 

On the Aerosense, the proposed methods are implemented with the Arm CMSIS library and the ADPCM directly works with quantized 15-bit variables while FFT-HPF is implemented with floating-point numbers. Since the CC2652P includes a floating-point unit there are no clear disadvantages in energy consumption and execution time between both approaches. Indeed, the energy savings of FFT-HPF are only worse than those of ADPCM by roughly 5\% and 1.5\% for small and large turbines, respectively. Combining with the better performance in terms of the time-domain and the frequency-domain distortions, FFT-HPF was selected for the final Aerosense version.
\subsection{Power Consumption and Lifetime Estimation}
\begin{figure}
    \centering
    \includegraphics[width=\columnwidth]{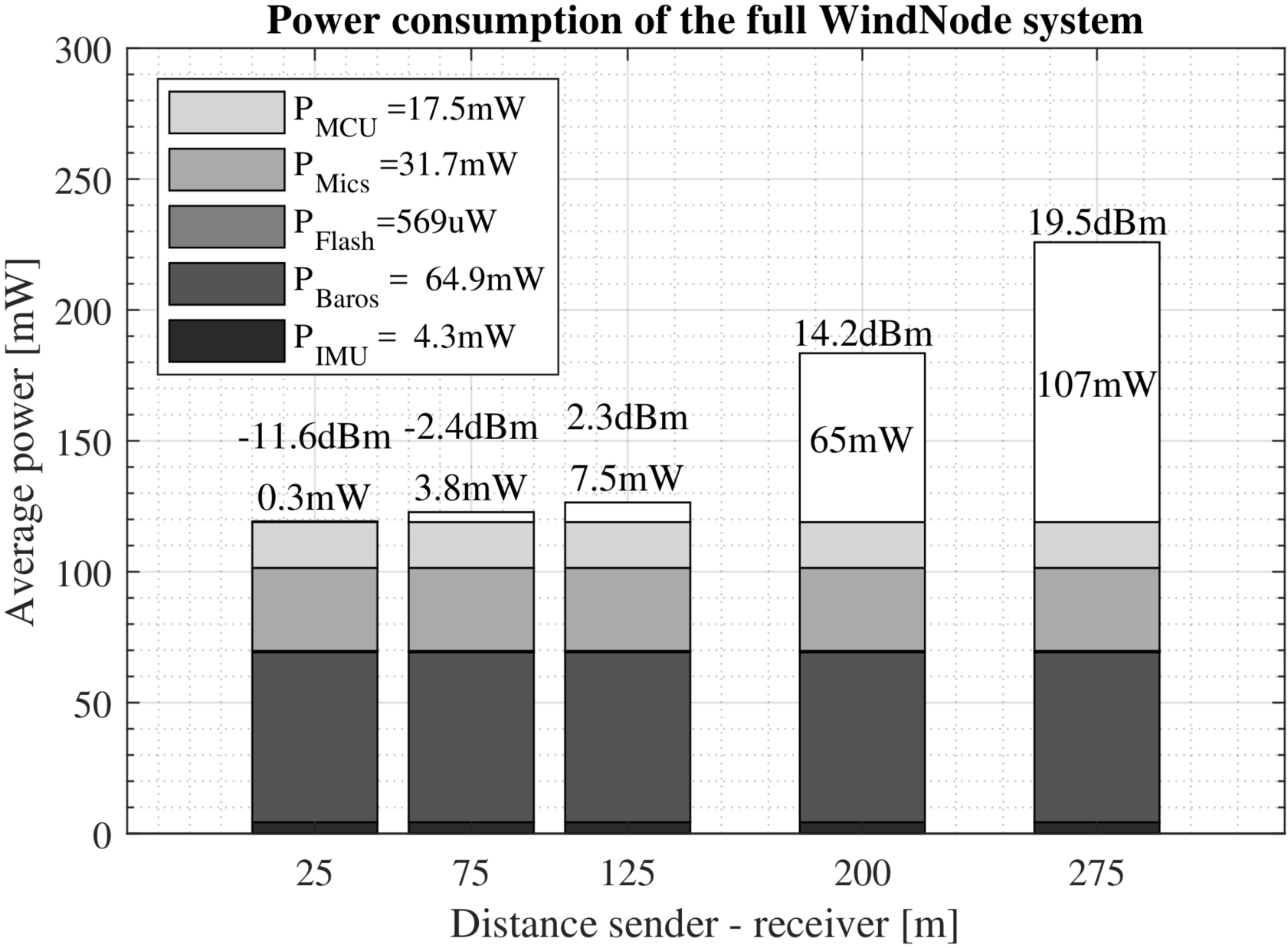}
    \caption{The Aerosense sensor node power consumption w.r.t. the distance from the base station. The measured power consumption is provided in detail for each sub-block, where the TX power expressed in dBm is correlated with the transmission range and it is automatically managed by the TPC.}
    \label{fig:powcons}
\end{figure}
%
%
Fig.~\ref{fig:powcons} shows the measured Aerosense power consumption at five measured distances. A detail of the power consumption of each block is provided, showing that barometers and the BLE impact power consumption the most, while the IMU and Flash operations are almost negligible. Hence we prove that storing and processing (compressing) the data on-board is a winning strategy for decreasing the average power consumption. 
Power consumption at different operational settings has been experimentally measured to estimate the Aerosense battery lifetime. We consider a 10-minute sampling interval every 120 minutes in our application scenario, with a corresponding duty cycle (DC) of 8.3\%. The system energy was measured outdoors at various distances from the base station, ranging from 25~m to 438~m. The sampling rates of the sensors were 16~kHz for the microphones, 100~Hz for the barometers, accelerometer, and gyroscope, and 12.5~Hz for the magnetometer, in addition to 1.2~kHz for differential pressure sensors. The base station was mounted at around 2~m height with a 17.5~dBi directional antenna.
The transmission power for a given distance was linearly interpolated from the measurements reported in Fig.~\ref{fig:powcons}, while lifetime and average power consumption are shown in Table~\ref{table:estlife}. It has been evaluated for two positions on an Aventa AV-7 turbine~\cite{aventa2021} with a rated power of 6.5~kW and a large 10~MW wind turbine~\cite{bak2013dtu}.
\begin{table}[t]
\centering
\caption{Average power consumption with two different wind turbines and the corresponding estimated lifetime}
\label{table:estlife}
\setlength{\tabcolsep}{3pt}
\begin{tabular}{l c c }
\hline\hline
\textbf{Description} & \textbf{Aventa AV-7} & \textbf{10~MW DTU} \\
 & \textbf{\cite{aventa2021}} & \textbf{\cite{bak2013dtu}} \\
\hline
Rot. Speed [rpm] & 40 & 10 \\
Hub Height [m] & 18 & 119 \\
Rotor Radius [m] & 6.5 & 79 \\
\hline
Average Power cons. (Active) [mW] & 142 & 135 \\
Average Power cons. (8.3\% DC) [mW] & 12 & 11 \\
Expected Lifetime [days] & 114 & 120 \\
\hline
Self-sustainability at $P_{median}$ & yes & yes \\
Self-sustainability at $P_{95\%}$ & no ($800~cm^2$)$^\star$ & no ($800~cm^2$)$^\star$ \\
\hline\hline
\multicolumn{3}{l}{$^\star$ Suggested solar panel size to reach self-sustainability}
\end{tabular}
\end{table}
To get realistic solar irradiance, global values from Leconfield~\cite{midas2019}, UK, were analyzed. 
From the data between 2014 and 2018, the harvestable solar power amounts to $P_{median} = 207~\mu W/cm^2$ and $P_{95\%} = 27.5~\mu W/cm^2$ for all but the 5\% worst days in winter.
Table~\ref{table:estlife} shows that the average power consumption in active mode is in the 140~mW range, whereas also considering periods of inactivity, it decreases to 10~mW. In this scenario, an ultra-low leakage current ($3~\mu A$ in total) enables a long battery lifetime, which exceeds 100 days of operation. However, thanks to the energy harvesting circuit, the self-sustainability is reached in $P_{median}$ conditions and can be reached for $P_{95\%}$ using an increased photovoltaic panel size (min. $800~cm^2$).

%% file: text/conclusion.tex
\section{Conclusion}
\label{sec:conclusion}
This paper presents Aerosense, a self-sustainable and long-range Bluetooth wireless sensor node for aerodynamic analysis and monitoring of wind turbines. 
The Aerosense can cover a wide range of different wind turbines in terms of size, rotor speed, and blade shape, aiming to collect helpful information to understand better the aerodynamics and aeroacoustic behavior of such renewable energy generators. We designed Aerosense around a BLE system-on-chip with enough bandwidth and coverage to transmit up to 438~m at high energy efficiency. The developed sensor node has been evaluated for functionality, range, and power consumption in a real application scenario. It has a low active power consumption of only 140~mW and thanks to the low sleep power consumption it can be duty-cycled at 8.3\% to reach self-sustainability with a $111~cm^2$ photovoltaic panel. Our system can support a sensor node with 40 barometers, 5 differential pressure sensors, and 10 microphones. In this configuration, the estimated hardware cost is below \SI{1000}{}\euro{}.
%
%

%% file: REV_1.bbl
\begin{thebibliography}{10}
\providecommand{\url}[1]{#1}
\csname url@samestyle\endcsname
\providecommand{\newblock}{\relax}
\providecommand{\bibinfo}[2]{#2}
\providecommand{\BIBentrySTDinterwordspacing}{\spaceskip=0pt\relax}
\providecommand{\BIBentryALTinterwordstretchfactor}{4}
\providecommand{\BIBentryALTinterwordspacing}{\spaceskip=\fontdimen2\font plus
\BIBentryALTinterwordstretchfactor\fontdimen3\font minus
  \fontdimen4\font\relax}
\providecommand{\BIBforeignlanguage}[2]{{%
\expandafter\ifx\csname l@#1\endcsname\relax
\typeout{** WARNING: IEEEtran.bst: No hyphenation pattern has been}%
\typeout{** loaded for the language `#1'. Using the pattern for}%
\typeout{** the default language instead.}%
\else
\language=\csname l@#1\endcsname
\fi
#2}}
\providecommand{\BIBdecl}{\relax}
\BIBdecl

\bibitem{suryakiran2020development}
M.~N.~S. Suryakiran, W.~Begum, R.~Sudhakar \emph{et~al.}, ``Development of wind
  energy technologies and their impact on environment: A review,''
  \emph{Advances in Smart Grid Technology}, pp. 51--62, 2020.

\bibitem{deshmukh2019wind}
S.~Deshmukh, S.~Bhattacharya, A.~Jain \emph{et~al.}, ``Wind turbine noise and
  its mitigation techniques: A review,'' \emph{Energy Procedia}, vol. 160, pp.
  633--640, 2019.

\bibitem{fischer2021windnode}
R.~Fischer, H.~Mueller, T.~Polonelli \emph{et~al.}, ``Windnode: A long-lasting
  and long-range bluetooth wireless sensor node for pressure and acoustic
  monitoring on wind turbines,'' in \emph{2021 4th IEEE International
  Conference on Industrial Cyber-Physical Systems (ICPS)}.\hskip 1em plus 0.5em
  minus 0.4em\relax IEEE, 2021, pp. 393--399.

\bibitem{di2021structural}
F.~Di~Nuzzo, D.~Brunelli, T.~Polonelli \emph{et~al.}, ``Structural health
  monitoring system with narrowband iot and mems sensors,'' \emph{IEEE Sensors
  Journal}, vol.~21, no.~14, pp. 16\,371--16\,380, 2021.

\bibitem{jin2020condition}
X.~Jin, Z.~Xu, and W.~Qiao, ``Condition monitoring of wind turbine generators
  using scada data analysis,'' \emph{IEEE Transactions on Sustainable Energy},
  vol.~12, no.~1, pp. 202--210, 2020.

\bibitem{madsen2010dan}
H.~Madsen, P.~Fuglsang, J.~Romblad \emph{et~al.}, ``The dan-aero mw
  experiments,'' in \emph{48th Aiaa Aerospace Sciences Meeting Including the
  New Horizons Forum and Aerospace Exposition}, 2010, p. 645.

\bibitem{mieloszyk2017application}
M.~Mieloszyk and W.~Ostachowicz, ``An application of structural health
  monitoring system based on fbg sensors to offshore wind turbine support
  structure model,'' \emph{Marine Structures}, vol.~51, pp. 65--86, 2017.

\bibitem{aagaard2010dan}
H.~{Aagaard Madsen}, C.~Bak, U.~{Schmidt Paulsen} \emph{et~al.},
  \emph{\BIBforeignlanguage{English}{The DAN-AERO MW experiments. Final
  report}}, ser. Denmark. Forskningscenter Risoe. Risoe-R.\hskip 1em plus 0.5em
  minus 0.4em\relax Danmarks Tekniske Universitet, Ris{\o} Nationallaboratoriet
  for B{\ae}redygtig Energi, 2010, no. 1726(EN).

\bibitem{maldonado2020using}
J.~Maldonado-Correa, S.~Mart{\'\i}n-Mart{\'\i}nez, E.~Artigao \emph{et~al.},
  ``Using scada data for wind turbine condition monitoring: A systematic
  literature review,'' \emph{Energies}, vol.~13, no.~12, p. 3132, 2020.

\bibitem{ahmad2021scopes}
I.~Ahmad, L.~M. Hee, A.~M. Abdelrhman \emph{et~al.}, ``Scopes, challenges and
  approaches of energy harvesting for wireless sensor nodes in machine
  condition monitoring systems: A review,'' \emph{Measurement}, vol. 183, no.
  109856, 2021.

\bibitem{ballerini2020nb}
M.~Ballerini, T.~Polonelli, D.~Brunelli \emph{et~al.}, ``Nb-iot versus lorawan:
  An experimental evaluation for industrial applications,'' \emph{IEEE
  Transactions on Industrial Informatics}, vol.~16, no.~12, pp. 7802--7811,
  2020.

\bibitem{fathima2021mems}
K.~M. Fathima, R.~S. Raj, K.~R. Prasad \emph{et~al.}, ``Mems multi sensor
  intelligent damage detection for wind turbines by using iot,'' in
  \emph{Journal of Physics: Conference Series}, vol. 1916, no. 012045.\hskip
  1em plus 0.5em minus 0.4em\relax IOP Publishing, 2021.

\bibitem{filipsky2017design}
J.~Filipsk{\`y}, J.~{\v{C}}{\'\i}{\v{z}}ek, F.~Wittmeier \emph{et~al.},
  ``Design and first test of the new synchronous 200 hz system for unsteady
  pressure field measurement,'' in \emph{FKFS Conference}.\hskip 1em plus 0.5em
  minus 0.4em\relax Springer, 2017, pp. 252--263.

\bibitem{karad2021efficient}
S.~Karad and R.~Thakur, ``Efficient monitoring and control of wind energy
  conversion systems using internet of things (iot): a comprehensive review,''
  \emph{Environment, Development and Sustainability}, vol.~23, 10 2021.

\bibitem{wondra2019wireless}
B.~Wondra, S.~Malek, M.~Botz \emph{et~al.}, ``Wireless high-resolution
  acceleration measurements for structural health monitoring of wind turbine
  towers,'' \emph{Data-Enabled Discovery and Applications}, vol.~3, no.~1,
  p.~4, 2019.

\bibitem{lu2019wind}
L.~Lu, Y.~He, T.~Wang \emph{et~al.}, ``Wind turbine planetary gearbox fault
  diagnosis based on self-powered wireless sensor and deep learning approach,''
  \emph{IEEE Access}, vol.~7, pp. 119\,430--119\,442, 2019.

\bibitem{esu2016feasibility}
O.~O. Esu, S.~D. Lloyd, J.~A. Flint \emph{et~al.}, ``Feasibility of a fully
  autonomous wireless monitoring system for a wind turbine blade,''
  \emph{Renewable Energy}, vol.~97, pp. 89--96, 2016.

\bibitem{madsen2019transition}
H.~A. Madsen, {\"O}.~S. {\"O}z{\c{c}}akmak, C.~Bak \emph{et~al.}, ``Transition
  characteristics measured on a 2mw 80m diameter wind turbine rotor in
  comparison with transition data from wind tunnel measurements,'' in
  \emph{AIAA Scitech 2019 Forum}, no. 0801, 2019.

\bibitem{aventa2021}
\BIBentryALTinterwordspacing
A.~AG. ([Online] 2021) Technische daten. (accessed: 01.02.2022). [Online].
  Available: \url{http://www.aventa.ch/technische-daten.html}
\BIBentrySTDinterwordspacing

\bibitem{bak2013dtu}
C.~Bak, F.~Zahle, R.~Bitsche \emph{et~al.}, ``The dtu 10-mw reference wind
  turbine,'' in \emph{Danish Wind Power Research 2013}, 2013.

\bibitem{jonkman2009definition}
J.~Jonkman, S.~Butterfield, W.~Musial \emph{et~al.}, ``Definition of a 5-mw
  reference wind turbine for offshore system development,'' National Renewable
  Energy Lab.(NREL), Golden, CO (United States), Tech. Rep., 2009.

\bibitem{chowdhury2020adaptive}
M.~R. Chowdhury, S.~Tripathi, and S.~De, ``Adaptive multivariate data
  compression in smart metering internet of things,'' \emph{IEEE Transactions
  on Industrial Informatics}, vol.~17, no.~2, pp. 1287--1297, 2020.

\bibitem{du2017compressed}
Z.~Du, X.~Chen, H.~Zhang \emph{et~al.}, ``Compressed-sensing-based periodic
  impulsive feature detection for wind turbine systems,'' \emph{IEEE
  Transactions on Industrial Informatics}, vol.~13, no.~6, pp. 2933--2945,
  2017.

\bibitem{solano2020optimal}
F.~Solano~Donado, ``On the optimal calculation of the rice coding parameter,''
  \emph{Algorithms}, vol.~13, no.~8, p. 181, 2020.

\bibitem{liaghati2018microcontroller}
A.~L. Liaghati, ``Microcontroller implementation of the biased dual-state
  dpcm,'' in \emph{2018 IEEE Aerospace Conference}.\hskip 1em plus 0.5em minus
  0.4em\relax IEEE, 2018, pp. 1--7.

\bibitem{midas2019}
\BIBentryALTinterwordspacing
C.~for Environmental Data~Analysis. (2019) Midas open: Uk hourly solar
  radiation data. (accessed: 01.02.2022). [Online]. Available:
  \url{http://web.archive.org/web/20080207010024/http://www.808multimedia.com/winnt/kernel.htm}
\BIBentrySTDinterwordspacing

\end{thebibliography}
